# Gel - Sol Transition of Thermoresponsive Poly(vinyl alcohol) Solution: Validation of the Universal Critical Scaling Relations

Tulika Bhattacharyya, Khushboo Suman,* [a)] and Yogesh M. Joshi*

Department of Chemical Engineering, Indian Institute of Technology Kanpur, Kanpur, Uttar Pradesh 208016, India

[a)] Present Address: Department of Chemical and Biomolecular Engineering, University of Delaware, Newark, Delaware 19716, USA

*Corresponding Authors, emails: ksuman@udel.edu and joshi@iitk.ac.in

**Abstract**

While undergoing gelation transition, a material passes through a distinctive state called the critical gel state. In the neighborhood of this critical gel state, how viscosity, equilibrium modulus, and relaxation times evolve are correlated by scaling relations, and their universality has been validated for materials undergoing the sol - gel transition. In this work, we extend this approach for the gel – sol transition of a thermoresponsive polymeric system of aqueous Poly(vinyl alcohol) (PVOH) gel that passes through the critical state upon increasing temperature. We observe that, in the neighborhood of the critical gel state, the equilibrium modulus and viscosity demonstrate a power law dependence on the relative distance from the critical state in terms of normalized temperature. Furthermore, the relaxation times in the gel and the sol state shows symmetric power law divergence near the critical state. The corresponding critical power law exponents and the dynamic critical exponents computed at the critical gel – sol transition state validate the scaling and hyperscaling relations originally proposed for the critical sol – gel transition very well. Remarkably, the dependence of complex viscosity on frequency at different temperatures shows a comprehensive mastercurve irrespective of the temperature ramp rate independently in the gel and the sol state. This observation demonstrates how the shape of relaxation time spectrum is independent of both the temperature as well as the ramp rate. Since sol – gel as well as the gel – sol transitions are opposite to each other, the applicability of the scaling relations validated in this work suggests broader symmetry associated with how the structure evolves around the critical state irrespective of the direction.

**Graphical Abstract**

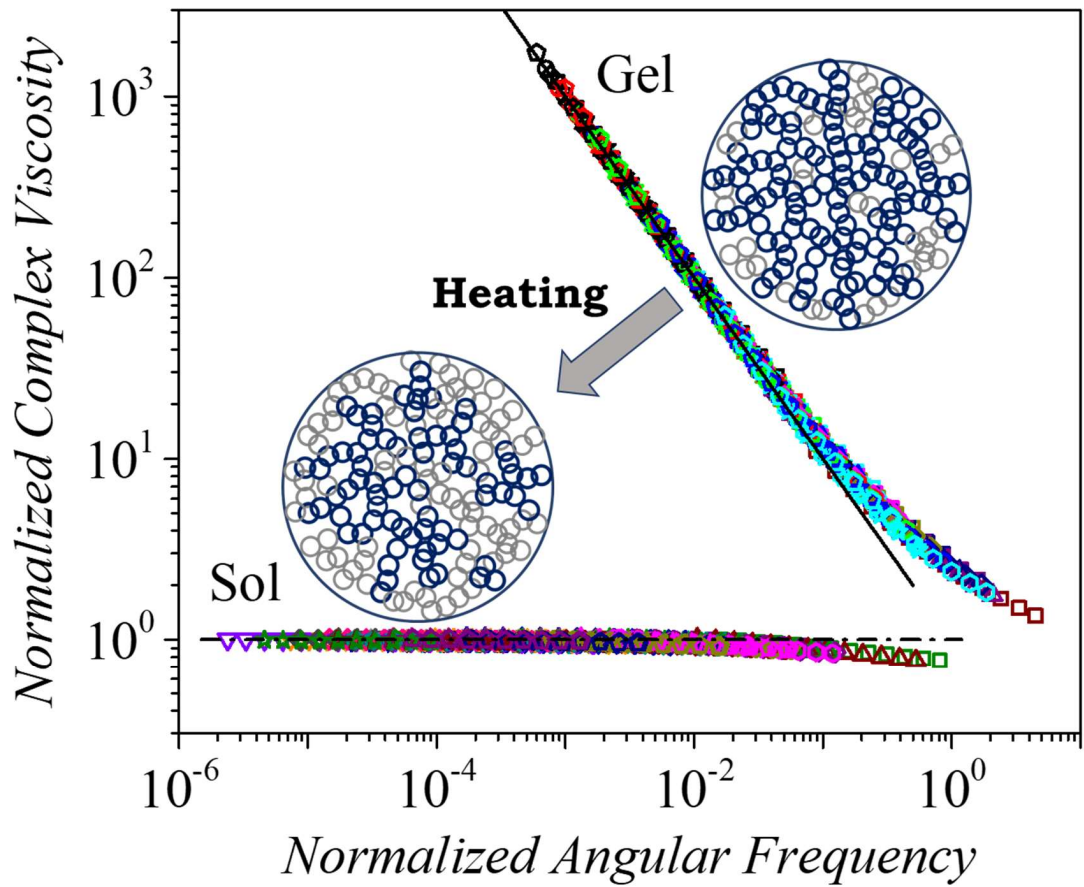

**Introduction:**

A wide variety of industrial systems undergo a liquid to gel transition owing to formation of a three-dimensional space spanning percolated network structure through chemical or physical crosslinking [1-5]. In chemical gels, the unsaturated polymer molecules, triggered by a chemical curing agent, forms a network structure through strong chemical (covalent) bonding [6]. On the other hand, in physical gels, relatively much weaker van der Waals interactions, hydrogen bonds, depletion interactions, or electrostatic bonds lead to sol - gel transition either spontaneously or with temperature change [7-10]. The critical sol - gel transition point is the distinctive state with the weakest space spanning percolated network structure [11, 12]. Owing to the self-similar network structure at the critical sol - gel transition state, the material properties show scale invariant behaviour. Accordingly, the relaxation modulus, compliance, and dynamic moduli exhibit a power law dependence with the critical relaxation exponent on their corresponding independent variable, time, or angular frequency [7, 12]. Furthermore, near the critical sol - gel transition state, the evolution of the viscoelastic properties like the zero shear viscosity, equilibrium modulus, and the characteristic relaxation time can also be expressed in terms of the extent of gelation in a power law fashion [13-16]. The critical exponent and the scaling exponents related to the zero shear viscosity, equilibrium modulus, and characteristic relaxation time demonstrate an exclusive interrelation through the hyperscaling laws. In the case of the sol – gel transition processes, the universality of the hyperscaling relations is established through theoretical [17-20] and experimental investigations on various chemical [6, 13, 15, 21, 22] and physical gels [23-26].

While the spontaneous formation of physical and chemical bonds is irreversible, temperature dependent hydrogen bonding can be reversed by altering the temperature field. Several aqueous polymeric solutions are observed to be thermoresponsive, where network structure during sol to gel and gel to sol transition can be tuned through temperature regulation. Owing to the extensive usage of such polymeric systems in tissue engineering, drug delivery, wound healing etc., it is necessary to understand the thermoresponsive network formation and dissociation of such materials [27-29]. Furthermore, in the food and beverage industry, knowledge of the complete formation and dissolution of gel structure of thermoresponsive gels is required to prevent blockage in the drying channels [30-32]. In clogged waxy crude oil pipelines, the flow is reestablished by applying DC electric field, which has been hypothesized to undergo network breakage, which is akin to gel - sol transition [33]. However, in the literature, not much attention has been devoted to the gel - sol transition compared to the sol - gel transition. Particularly it is important to understand how various scaling relations depend on each other and on the extent of crosslinking as the material approaches the critical state from a dense network gel state. The objective of this study is to explore the applicability of the universal hyperscaling laws on the gel to sol transition of a thermoresponsive polymeric gel.

The critical gel transition state is the unique state at which the material properties show a universal simplistic combination of liquids and solids. On the one hand, the steady shear viscosity at this state is infinite, like a solid, whereas on the other hand, it exhibits zero equilibrium modulus like a liquid [12]. The relaxation modes of the system at the critical gel state are coupled in such a way that they exhibit a power law distribution. Consequently, the linear stress relaxation modulus $G(t)$ and the continuous relaxation time spectra $H(\tau)$ decays with corresponding independent variable in a power law fashion, where the coefficient is critical relaxation exponent $n_c$, given by [12, 34]:

$$G(t) = St^{-n_c} \quad \text{for, } \tau_0 \leq t < \infty, \tag{1}$$

$$H(\tau) = \frac{S}{\Gamma(n_c)} \tau^{-n_c} \quad \text{for } \tau_0 \leq t < \infty \tag{2}$$

Here, $\Gamma(n_c)$ is the Euler gamma function of the critical relaxation exponent $n_c$ and $S = G_0 \tau_0^{n_c}$ is the gel strength [35]. In the expression of the gel strength, $G_0$ is the modulus associated with a fully developed network structure and $\tau_0$ is the relaxation time associated with the primitive link. The critical relaxation exponent $n_c$ varies in the range of 0 to 1, where $n_c \to 0$ denotes a constant modulus as observed in Hookean solids, while $n_c \to 1$ represents the viscous limit. The gel strength $S$ is inversely related to $n_c$ and takes the unit of Pa. $s^{n_c}$. Consequently, at the critical gel state, the elastic $(G')$ and viscous $(G'')$ moduli show an identical power law dependence on angular frequency $(\omega)$ with coefficient $n_c$ given by [36, 37]:

$$G' = G'' \cot\left(\frac{n_c \pi}{2}\right) = \frac{\pi S}{2\Gamma(n_c)\sin\left(\frac{n_c \pi}{2}\right)} \omega^{n_c}. \tag{3}$$

The loss tangent, defined as $\tan \delta = \tan\left(\frac{n_c \pi}{2}\right) = \frac{G''}{G'}$, therefore, decreases (or increases) with an increase in $\omega$ in the sol (or gel) state, but remains independent of $\omega$ at the critical gel state. In the seminal work by Winter and Chambon [36], the unique point of $\tan \delta$ being independent of $\omega$ is characterized as the rheological signature of sol - gel transition. The Winter- Chambon criterion has been verified for various sol - gel transition processes with chemical and physical crosslinking [6, 12, 23, 36-40]. The unique power law dependence of material properties at the critical gel transition state is indicative of a fractal-like self-similar morphology [41]. The self- similar structure over a wide lengthscale, in turn, results in a scale invariant relaxation behavior. Interestingly, the power law rheology exhibited at the critical sol - gel transition state is governed by only a single parameter, the critical relaxation exponent $n_c$. Furthermore, for a percolated network structure the knowledge of $n_c$ also leads to information about the fractal dimension ($f_d$). Muthukumar [41] proposed a relationship between the critical relaxation exponent, $n_c$ and the fractal dimension $f_d$, assuming that the network structure at the critical gel transition state is analogous to a polymer melt system with random branches having completely screened excluded volume effect:

$$f_d = \frac{5(2n_c - 3)}{2(n_c - 3)}. \qquad (4)$$

This suggests that the knowledge of $n_c$ provides significant insight in understanding the hierarchical three-dimensional network structure at the critical sol - gel transition state.

In a typical sol to gel transition process, it is customary to measure the network connectivity with an independent variable $p$ [12]. The definition of $p$ strictly depends on the type of crosslinking reaction. In case of a sol - gel transition of an unsaturated polymeric system, governed by chemical crosslinking reaction, $p$ is defined as the ratio of the number of crosslinks at that stage to the total possible number of crosslinks. The degree of crosslinking $p$, thus, varies in the range of 0 to 1 during the sol - gel transition process. At the critical sol - gel transition state, the degree of crosslinking is defined as $p = p_c$, however, the exact magnitude of $p_c$ specifically depends on the properties of the evolving system. Consequently, $p - p_c$ becomes the measure of the distance from the critical sol - gel transition state and the normalized distance from both the pre-gel and post-gel state can be expressed as $\Delta \tilde{p} = |p - p_c|/p_c = \Delta p/p_c$. As the transition proceeds in the pre-gel state, the formation of more chemical bonds leads to the growth of the cluster size. Adam *et al.* [16] reported that for a chemically crosslinking polymeric system, the largest cluster size ($\xi$) in the sol state diverges in a power law fashion with $\Delta \tilde{p}$ as:

$$\xi \sim \Delta \tilde{p}^{-\alpha}. \tag{5}$$

Owing to the continuously growing size of the largest cluster, the zero shear viscosity, $\eta_0$ and the dominant relaxation time, $\tau_{max,S}$ of the sol state increases. In the neighborhood of the critical sol - gel transition state, both $\eta_0$ and the corresponding $\tau_{max,S}$ of the pre-gel state diverges due to the diverging cluster size [20, 38, 42]:

$$\eta_0 \sim \Delta \tilde{p}^{-s} \tag{6}$$

$$\tau_{max,S} \sim \Delta \tilde{p}^{-v_S}. \tag{7}$$

Such growth of the largest cluster in turn leads to the transition from the sol to gel state through a critical gel point. The equilibrium modulus, $G_e$ is zero at the critical gel state, and gradually increases with subsequent crosslinking. Participation of more monomeric units in the cluster decreases the number of relaxable components, and the dominant relaxation time, $\tau_{max,G}$ continuously decreases as the system moves away from the critical sol - gel transition state. In the post-gel state, $G_e$ and $\tau_{max,G}$ evolve as [20, 38, 42]:

$$G_e \sim \Delta \tilde{p}^z \tag{8}$$

$$\tau_{max,G} \sim \Delta \tilde{p}^{-v_G}. \tag{9}$$

The scaling exponents are positive constants and distinct to the properties of the evolving system. The critical scaling exponents in the pre-gel and post-gel states are independently related to the critical relaxation exponent $n_c$ as [38]:

$$v_S = \frac{s}{1-n_c} \quad \text{and} \tag{10}$$

$$v_G = \frac{z}{n_c}. \tag{11}$$

The scaling relations Eqs. (6) to (9) hold true only near the critical gel state, where the material properties are strongly governed by the specific nature of the critical gel state [12].

Accordingly, the interrelations of the critical exponents $s$, $z$, $v_S$, and $v_G$ with the critical relaxation exponent $n_c$ through Eqs. (10) and (11) are dictated by the linear viscoelastic principles and, therefore, are widely reported to hold true. However, the critical exponents in the pre-gel and post-gel may not be strictly correlated. A special case arises when the longest relaxation time diverges symmetrically around the critical gel point such that [38]:

$$v_S = v_G. \qquad (12)$$

Such symmetric divergence of the relaxation time leads to a profound hyperscaling relationship given by [38]:

$$n_c = \frac{z}{s+z}. \qquad (13)$$

The hyperscaling relationship has also been derived theoretically [17-19], suggesting that the longest relaxation times diverge symmetrically on both sides of the critical gel state. The universality of the scaling and hyperscaling laws has been established for various chemical gels [15, 16, 43-47]. The divergence of $\eta_0$ in the pre gel state [Eq. (6)] and growth of $G_e$ in the post gel state [Eq. (8)] have been well established for several chemical gels like radical copolymerization of styrene and meta divinyl benzene [16], polycondensation of hexamethyl di—isocyanate and triol [16], polycondensation of polyurethanes [48] and PEG-Heparin hydrogels [49]. The hyperscaling law relating the scaling exponents $n_c$, $s$, and $z$ given by Eq. (13) have been verified for hydrolysis and condensation reaction of tetrathoxysilane [15], gelation of randomly branched polyster [43], and cupric cation induced gelation of aqueous alginate solution [44]. The hyperscaling law has also been established by experimentally obtaining the symmetric divergence of the relaxation time [Eq. (12)] for chemically crosslinking pectin hydrogels [45], peptide hydrogels and polyacrylamide gels [46, 47].

In case of physically crosslinking systems, the sol to gel transition takes place either spontaneously [39, 50-56] or with temperature change [25, 40, 57-64]. Here, since $\Delta\tilde{p}$ is not easily measurable in terms of connectivity, a normalized relative distance from the critical gel point, $\varepsilon = |X - X_c|/X_c$ can be defined in terms of a measurable reaction coordinate [12]. The variable $X$ can be time ($t$), temperature ($T$) or concentration ($c$) depending on the driving cause for gelation [23, 65, 66]. Studies [23, 65-67] have shown that Eqs. (6) to (9) also holds for a physical gel, when $\Delta \tilde{p}$ is replaced by $\varepsilon$. To some extent, the applicability of the hyperscaling laws have also been explored for various time and temperature dependent physical gels [25, 65-71]; however, the measure of the extent of gelation, $\varepsilon$ for such gels have not been conclusively established. For a temperature dependent gelation of gelatin [67, 68], the parameters $n_c$, $s$, $z$ and their interdependence [Eq. (13)] has been verified considering the measure of helix fraction as the extent of reaction. The hyperscaling law has also been established by assuming gelation concentration as $\varepsilon$ for temperature dependent gels of Cellulose [69], Methylcellulose [70], Gellan Gum [71] and Kappa Carrageenan [25]. Cho and Heuzey [65] have verified the hyperscaling law considering gelation time as $X$ for isothermal gelation experiments on Chitosan gel. Tan et al. [66] confirmed the applicability of the scaling laws for Polyacrylonitrile/ DMSO solution, assuming both aging time and gelation concentration as $\varepsilon$. Suman and Joshi [23] obtained the individual scaling exponents $n_c$, $s$, $z$, $v_S$, and $v_G$ considering $X = T$ for a temperature dependent polymeric gel and $X = t$ for a spontaneously evolving colloidal gel. They experimentally demonstrated the symmetric divergence for physical gels [Eq. (12)] and thus verified the universality of the hyperscaling law.

While the sol - gel transition is routinely studied in the literature, understanding of the gel - sol transition is still in its infancy. The gel to sol transition involves the dissociation of the three-dimensional percolated network structure on reversal of the gel forming variable. Certainly, under practical purposes, only the thermoresponsive polymeric gels undergo a gel to sol transition on changing the direction of application of temperature field [40, 58, 59, 64]. It is to be noted here that while the network structure of any gel can be deformed by shear rejuvenation, the process does not qualify as a gel to sol transition as applied stress/ strain is much higher than the linear viscoelastic limit for the gel. For the sol (liquid) - gel transition of a colloidal suspension, Jatav and Joshi [72] reported that while a freshly prepared sample demonstrates a critical gel state according to the Winter- Chambon criteria [36], the same is not observed for a shear rejuvenated system. Along similar lines, the dissociation of the network structure due to the degradation of gels, either spontaneously or through a change in environment, may also not always qualify as gel to sol transition. In case of hydrogenated castor oil gels, formed due to an osmotic pressure gradient, Wehrman et al. [73] reported that even after complete degradation, the system does not undergo a complete gel to sol transition. On the other hand, microrheology measurements of equilibrium compliance and the longest relaxation time during the degradation reaction of synthetic hydrogels indicated gel - sol transition [74, 75]. Several polymeric systems like Cellulose [69], Methylcellulose [70], Gellan Gum [71], Gelatin [76], Poly(vinyl alcohol) [40] and various types of Carrageenans [24, 25, 64] etc., undergo a sol to gel/ gel to sol transition through hydrogen bonding as a function of decreasing/ increasing temperature. While the universality of the scaling laws has been established for the sol - gel transition processes of various polymeric systems, the gel to sol transition process is yet to be investigated.

The present study focuses on understanding the thermoresponsive transition of an aqueous polymeric system upon heating. Firstly, we study how the linear viscoelastic properties evolve during the gel - sol transition. Subsequently, we investigate the applicability of the scaling relations proposed for sol - gel transition on the gel to sol transition process. To summarize, we analyze the interrelations of the scaling exponents related to the gel - sol transition and highlight the similarity of the thermoresponsive sol - gel transition with the same.

**Materials and Methods:**

In this work, we use an aqueous solution of Poly(vinyl alcohol) (PVOH), a thermoresponsive system that undergoes gelation transition with change in temperature. PVOH is a semi-crystalline, biocompatible, and nontoxic polymer with high mechanical strength. With a decrease in temperature, the aqueous solution of PVOH undergoes sol - gel transition owing to network structure formation through hydrogen bonding. On reversal of the temperature field, the hydrogen bonds start to dissociate, leading to a gel to sol transition. PVOH of molecular weight 89,000-98,000 g/mol (>99% hydrolyzed) is procured from Sigma Aldrich. We prepare a 12 wt.% aqueous solution in Millipore water of resistivity 18MΩ. cm. A predetermined amount of polymer (oven dried at a constant temperature of 120°C for 12 h) is added to 200ml of Millipore water, and the solution is continuously stirred in an IKA C-MAG HS7 magnetic stirrer at 400 rpm for 8 h at 80°C. The mixing is carried out in a closed container to prevent water losses due to evaporation. The freshly prepared aqueous solution of PVOH is then cooled to the ambient condition of 30°C and stored for 4 h before any experiment.

The rheological measurements are performed in a Dynamic Hybrid Rheometer 3 rheometer by TA instruments. The temperature is maintained with a Peltier plate temperature control assembly. We use a smooth concentric cylinder geometry with a diameter= 28.14 mm and a gap thickness = 1.2 mm. The aqueous PVOH solution is put in the shear cell at 30°C and subjected to a temperature down ramp at a ramp rate ($k$) of 0.05 °C/min. The gel is further consolidated by equilibration at 0°C for 6 h. Subsequently, we apply a temperature up ramp to the PVOH gel at different rates, from $k = +0.05$ °C/min to $k = +0.4$ °C/min. During application of the temperature ramp in all the steps, the system is probed with a cyclic frequency sweep with angular frequency varying from $\omega$=0.5 rad/s to $\omega$=25 rad/s at a constant magnitude of stress, $\sigma = 1$ Pa, which is verified to be within the linear viscoelastic regime. Each frequency sweep measurement takes less than 100 s for completion. Since over this timescale, temperature undergoes a very small change, the evolution of the PVOH gel during the frequency sweep can be safely considered to be negligible. For all the rheological measurements, we cover the free surface of the sample with a low-density silicone oil to minimize evaporation losses. Each measurement has been repeated at least twice to confirm the reproducibility of data. To distinguish the two transition processes, we represent the sol and gel states of sol - gel transition as pre-gel (prior to the critical sol - gel transition state) and post-gel (post the critical sol - gel transition state) respectively. Furthermore, we denote the data corresponding to sol to gel transition with subscript $-SG$, and gel to sol transition with subscript $-GS$.

In a typical procedure, PVOH solution is cooled from 30°C to 0°C by applying a temperature ramp of $k = -0.05$ °C/min. PVOH solution is in the sol state at 30°C. As temperature is decreased, it undergoes sol - gel transition. We monitor this process by applying succession of frequency ramps in the linear viscoelastic regime. We report various rheological characteristic features of the sol - gel transition in the supplementary information, wherein Figure S1 describes how dynamic moduli and tan$\delta$ evolve as PVOH undergoes the gelation transition. Various scaling and hyperscaling relations mentioned in the introduction have been experimentally validated for the sol - gel transition [23].

**Results and Discussions:**

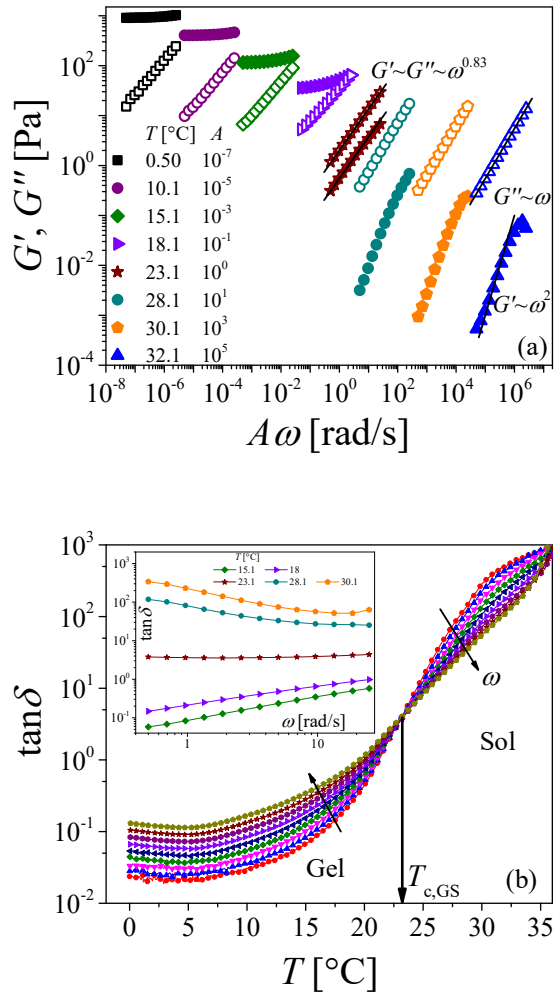

**Figure. 1:** (a) Evolution of $G'$ (closed symbol) and $G''$ (open symbol) plotted as a function of $\omega$ at different temperatures for gel - sol transition upon heating from 0°C at a ramp rate of $k = +0.05$ °C/min. The power law fitting lines are shown at the critical gel - sol transition state and the sol state. For better clarity, the data has been shifted horizontally with shift factor $A$, and the magnitude of $A$ at any temperature is mentioned in the legends. (b) The corresponding evolution of $\tan\delta$ as a function of temperature is shown for gel - sol transition. The slanted arrows show the direction of increasing $\omega$. In the inset, we plot the evolution of $\tan\delta$ as a function of $\omega$ at different temperatures. The critical gel - sol transition state, where $\tan\delta$ is independent of $\omega$, at temperature $T_{c,GS} = 23.1$°C is marked with a solid arrow.

In order to characterize the gel - sol behavior, the PVOH gel at 0°C is subjected to a temperature up ramp at a rate of $k = +0.05$ °C/min along with cyclic frequency sweep. Figure 1(a) shows the corresponding dependence of elastic ($G'$) and viscous ($G''$) moduli on angular frequency ($\omega$) with increasing temperatures. For better clarity, the horizontal axis of $\omega$ has been shifted with shift factor $A$. At lower temperatures, the gel shows a solid-like behavior, where $G'$ remains nearly independent of $\omega$. With an increase in temperature, $G'$ decreases, but its dependence on $\omega$ starts to grow. On the other hand, $G''$ shows a power law dependence on $\omega$. Interestingly, the corresponding power law coefficients seem to have a weak dependence on temperature through the value of $G''$ (at constant $\omega$) that decreases with an increase in temperature. Eventually both $G'$ and $G''$ exhibit an identical power law dependence on $\omega$ ($G' \sim G'' \sim \omega^{0.83}$) at temperature, $T_{c,GS} = 23.1$°C, which represents the critical gel state. The identical values of the power-law exponent for dependence of $G'$ and $G''$ on $\omega$ confirms the existence of the weakest self-similar percolated network structure at the critical gel - sol transition state [36, 37]. With further increase in temperature, the power law dependence of $G'$ and $G''$ on $\omega$ increases. At higher temperature, the system demonstrates signatures of the sol state where, $G' \sim \omega^2$ and $G'' \sim \omega$. The corresponding evolution of $\tan \delta$ as a function of temperature at different $\omega$ is shown in Figure 1(b). In the inset of Figure 1(b), we plot $\tan \delta$ as a function of $\omega$ at different temperatures. At lower temperatures, since $G' \gg G''$, we observe $\tan \delta \ll 1$ while the values of $\tan \delta$ increase with increase in $\omega$. At the point of critical gel - sol transition state, $\tan \delta$ is independent of $\omega$ as shown in the inset of Figure 1(b). Accordingly, the isofrequency $\tan \delta$ curves merge at a single point, as marked with a solid arrow in Figure 1(b). This leads to information of the critical relaxation exponent using the relation $\tan \delta = \tan \left( n_{c,GS} \frac{\pi}{2} \right)$, thus indicating $n_{c,GS} = 0.83$, which precisely agrees with the frequency dependence of $G'$ and $G''$ on $\omega$ at $T_{c,GS} = 23.1$°C. On further increase in temperature, in the sol state, the dependence of $\tan \delta$ on $\omega$ reverses and $\tan \delta \gg 1$, as $G'' \gg G'$. Therefore, PVOH solution strongly exhibits all the rheological signatures of gel - sol transition upon heating.

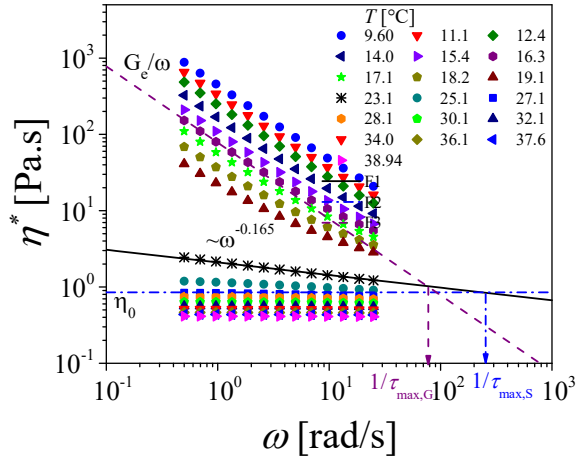

**Figure 2:** Evolution of $|\eta^*|$ as a function of $\omega$ at different temperatures during gel to sol transition at a ramp rate of $k = +0.05$ °C/min. The power law dependence of $|\eta^*| \sim \omega^{(-1+n_{c,GS})}$ at the gel to sol transition state with $n_{c,GS} - 1 = -0.165$ is shown with a solid line. The low frequency asymptote in the gel state, $G_e(t,\varepsilon)/\omega = \lim_{\omega \to 0}|\eta^*|$, where $G_e$ is the equilibrium modulus, is shown with a dashed line. The low frequency asymptote in the sol state, $|\eta^*| = \eta_0$, where $\eta_0$ is the zero shear viscosity, is shown with a dash-dot line. The intersection point of the low frequency asymptotes with gel to sol transition state is marked, and the reciprocal of the corresponding $\omega$ is the characteristic relaxation time in the gel ($\tau_{max,G}$) and the sol ($\tau_{max,S}$) state respectively.

The cyclic frequency sweep measurements during the gel - sol transition also provides us with information on complex viscosity: $|\eta^*(\omega, p)| = \sqrt{G'^2 + G''^2}/\omega$. In Figure 1, the oscillatory measurements establish 0°C as the gel region where the microstructure of PVOH gel is proposed to consist of a dense network formed by the polymer segments connected through inter/intrachain hydrogen bonding. The interconnected polymer segments constitute the microcrystalline domains, which eventually lead to junction knots for network structure formation. On subjecting this consolidated gel structure to heating, the hydrogen bonds start to break, and the system undergoes a gel to sol transition. In Figure 2, we plot how the corresponding $|\eta^*|$ evolves as a function of $\omega$ during the increase in temperature. In the limit of low temperatures $|\eta^*|$ decreases with an increase in frequency.

With increase in temperature, the dependence of $|\eta^*|$ on $\omega$ decreases as the network structure of the gel gradually dissociates. More specifically, at temperatures lower than the critical gel - sol transition temperature ($0 < T < T_{c,GS}$), $|\eta^*|$ diverges in the limit of small $\omega$. In the gel state, a solid-like response is reported in the low $\omega$ regime, wherein $|\eta^*|$ shows an inverse dependence on $\omega$ given by: $\lim_{\omega \to 0} |\eta^*| = G_e(t,\varepsilon)/\omega$ [38], leading to the estimation of the equilibrium modulus, $G_e$ of a consolidated gel. Consequently, we construct a low $\omega$ asymptote having slope of $-1$, as shown by dashed line in Figure 2 which leads to the quantification of $G_e$. Winter [38] report that in the post-gel state in the limit of low $\omega$ very few datapoints of $|\eta^*|$ versus $\omega$ plot follow an inverse dependence on $\omega$, and this dependence diminishes as the system approaches the critical sol - gel transition state. Accordingly, with increase in temperature, as the system approaches the critical gel - sol transition state, the number of datapoints following the inverse dependence on $\omega$ gradually decreases. At the critical gel - sol transition state, $|\eta^*|$ follows a power law dependence on $\omega$, similar to what has been reported at the critical sol - gel transition and is given by [38]:

$$\eta^*(\omega,\varepsilon) = aS(i\omega)^{(-1+n_{c,GS})} \quad \text{or,}$$

$$|\eta^*(\omega,\varepsilon)| = aS\omega^{(-1+n_{c,GS})}. \qquad (14)$$

This dependence of $|\eta^*|$ on $\omega$ at the critical gel-sol transition temperature $T_{c,GS} = 23.1°C$, has been shown in Figure 2 with a solid line. The validation of Eq. (3) and (14) characterizes the existence of the weakest self-similar space spanning percolated network structure at the critical gel - sol transition temperature. With further increase in temperature, more junction knots dissociate as hydrogen bonds dissolve. In the sol state, $|\eta^*|$ shows a plateau in the low $\omega$ regime as observed for the pre-gel state. This facilitates the estimation of the zero shear viscosity: $\eta_0 = \lim_{\omega \to 0} |\eta^*|$. The low frequency asymptote to obtain $\eta_0$ in the sol state is marked with the dash-dot line in Figure 2. The plateau value representing $\eta_0$ decreases with an increase in temperature as the network structure continues to weaken. In the limit of high frequency, Winter [38] reported that both in sol and gel states, $|\eta^*|$ follows a power law dependence on $\omega$ denoted by Eq. (14). The magnitude of $\omega$ at which $|\eta^*|$ starts to follow the power law dependence decreases as the gel approaches the critical gel - sol transition state and increases as it moves further into the sol state.

Interestingly, the inverse of the $\omega$ associated with the intersection point of the power law dependence at the critical sol - gel transition state, with low $\omega$ asymptotes in the pre-gel and post-gel state, leads to the estimation of their corresponding dominant relaxation times. Accordingly, we obtain the dominant relaxation time at the gel state, $\tau_{max,G}$ (indicated by the dashed arrow in Figure 2) and the sol state, $\tau_{max,S}$ (indicated by the dash-dot arrow in Figure 2) during the gel to sol transition with an increase in temperature. Such construction shown in Figure 2, therefore, facilitates the estimation of $G_e$, $\eta_0$, $\tau_{max,G}$ and $\tau_{max,S}$ at different temperatures during the gel to sol transition. It is to be noted that the construction of low frequency asymptotes and estimation of parameters in the sol and gel region for the PVOH system has been performed here with oscillatory shear measurements, as proposed for the pre-gel and post-gel state of a polymeric system [38]. The evolution of the linear viscoelastic properties can also be measured through steady shear measurements. However, owing to the fragile network structure of the critical gel, this measurement technique has been observed to be destructive [67].

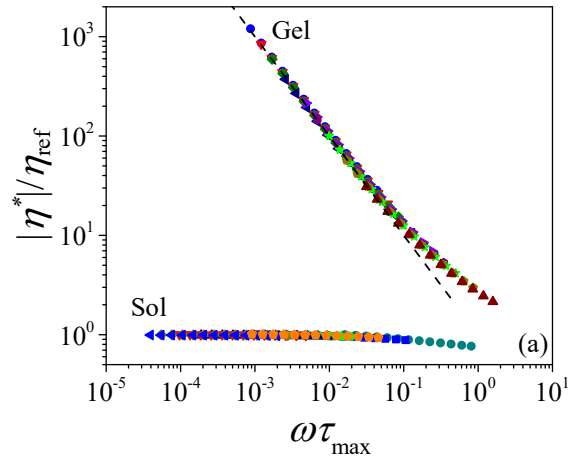

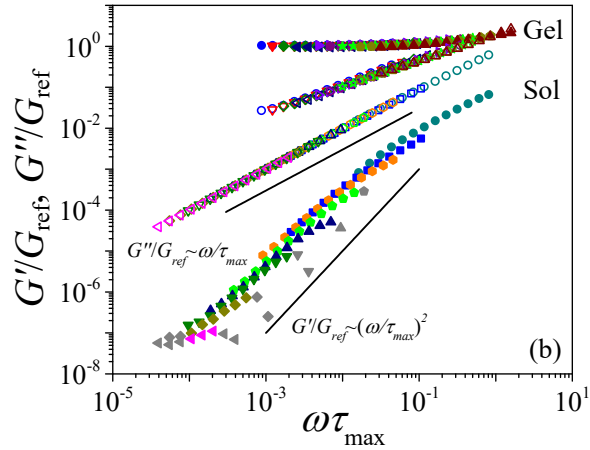

**Figure 3:** (a) Normalized $|\eta^*|$ plotted as a function of normalized $\omega$ at different temperatures during gel - sol transition at a ramp rate of $k = +0.05$ °C/min. The normalization factor for ordinate in the gel state is $\eta_{ref} = G_e \tau_{max,G}$ and in the sol state is $\eta_{ref} = \eta_0$. The normalization factor for the abscissa is the characteristic relaxation time in case of gel and sol states, $\tau_{max,G}$ and $\tau_{max,S}$. The normalized low frequency asymptote in the gel state is shown with a dashed line of slope $-1$. (b) Normalized $G'$ and $G''$ plotted as a function of normalized $\omega$ at different temperatures during gel - sol transition at a ramp rate of $k = +0.05$ °C/min. The normalization factor of $G'$ and $G''$ for ordinate in the gel state is $G_{ref} = G_e$ and in the sol state is $G_{ref} = \eta_0/\tau_{max,S}$. The normalization factor for the abscissa is the characteristic relaxation time in case of the gel ($\tau_{max,G}$) and the sol ($\tau_{max,S}$) state. In the sol state, the system shows signatures of a liquid, where $G' \sim \omega^2$ and $G'' \sim \omega$, as shown with solid lines.

With the estimation of $G_e$, $\eta_0$, $\tau_{max,G}$ and $\tau_{max,S}$ from the low $\omega$ asymptote construction, we attempt to rescale the curves associated with the complex viscosity and the dynamic moduli as a function of $\omega$ using the corresponding estimated variables at a specific temperature. In Figure 3(a) we plot $|\eta^*|$ normalized with a reference parameter $\eta_{ref}$ as a function of $\omega\tau_{max}$. The corresponding normalization parameter for the gel state is $\eta_{ref} = G_e\tau_{max,G}$ and $\tau_{max} = \tau_{max,G}$, while for the sol state, $\eta_{ref} = \eta_0$ and $\tau_{max} = \tau_{max,S}$.

Interestingly, the plot with normalized abscissa and ordinate shows excellent superpositions independently in the sol and gel domains. The superposed data in the regime of low frequency shows an inverse dependence on $\omega$ in the gel state but remains constant in the sol state. On the other hand, at higher $\omega$ the data shows a power law behaviour both in gel and sol states. Such superposition with normalized $|\eta^*|$ and $\omega$ with accurate estimation of $G_e$, $\eta_0$, $\tau_{max,G}$ and $\tau_{max,S}$ have also been reported previously for the sol – gel transition with physical or chemical crosslinking [23, 38]. Furthermore, in Figure 3(b), we replot Figure 1(a), the dependence of $G'$ and $G''$ on $\omega$, normalizing the ordinate and abscissa with corresponding reference parameters $G_{ref}$ and $\tau_{max}$ respectively. The normalization parameter for the gel state is $G_{ref} = G_e$ and $\tau_{max} = \tau_{max,G}$, while for the sol state is $G_{ref} = \eta_0/\tau_{max,S}$ and $\tau_{max} = \tau_{max,S}$. Accordingly, we observe the plots with normalized abscissa and ordinate to superpose independently in the sol and gel domains. The normalized data in the sol domain demonstrates a distinct dependence of $G'/G_{ref} \sim (\omega\tau_{max})^2$ and $G''/G_{ref} \sim \omega\tau_{max}$, as indicated with the solid lines, thus suggesting a liquid state. In the gel state, on the other hand, $G'/G_{ref}$ shows weak dependence on $\omega$ while $G''/G_{ref}$ shows a power law dependence on $\omega$. The existence of superposition plots shown in Figure 3 in turn confirm the accurate estimation of the $G_e$, $\eta_0$, $\tau_{max,G}$ and $\tau_{max,S}$ during the gel - sol transition. Furthermore, the superposition obtained in Figure 3 is a distinctive one, as the shift factors are not just arbitrary values but actually represent the viscoelastic parameters specific to the material and the gel - sol transition. The two separate superpositions in the gel and sol states indicate that the shape of the relaxation time spectra is conserved independently in the gel and the sol state. It is necessary to note that the critical gel state is not included in the superposed curve as the largest relaxation timescale associated with the space spanning network at the critical gel state becomes infinite.

For a chemically crosslinking sol - gel transition, zero shear viscosity $\eta_0$, equilibrium modulus $G_e$, and the longest relaxation times in the sol ($\tau_{max,S}$) and the gel ($\tau_{max,G}$) state scales with $\Delta\tilde{p}$. On the other hand, for sol - gel transition processes with physical crosslinks, the extent of reaction is measured in terms of the variable that drives the transition: time, temperature or concentration. In this work, we measure the evolution of viscoelastic properties during the gel - sol transition as a function of relative distance from the critical

gel - sol transition temperature, $\varepsilon = |T - T_{c,GS}|/T_{c,GS}$, where $T_{c,GS}$ is the critical gel - sol transition temperature. A lower value of $\varepsilon$ corresponds to a temperature closer to the critical gel state, and a higher value corresponds to a state far away from the critical gel state. This approach is in accordance with Suman and Joshi [23], who studied the nonisothermal sol - gel transition of PVOH. In Figures 4 (a) and (c), we plot the evolution of $G_e$ and $\tau_{max,G}$ in the gel state, and $\eta_0$ and $\tau_{max,S}$ in the sol state as a function of $\varepsilon = |T - T_{c,GS}|/T_{c,GS}$. The methodology to obtain these parameters is shown pictorially in Figure 2. Gels show solid-like nature and are characterized by the presence of $G_e$, as measured by the low $\omega$ asymptote. An increase in the temperature of the consolidated PVOH gel, equivalent to a decrease in $\varepsilon$, results in a decrease in $G_e$ as the network structure dissociates. The longest relaxation time, $\tau_{max,G}$ is associated with the relaxable components of the gel, i.e., the free PVOH chains which are not participating in the network structure [12]. Accordingly, $\tau_{max,G}$ increases with decrease in $\varepsilon$ as more free chains become available due to network dissociation. The Figures 4 (a) and (c) illustrate that in the vicinity of the critical gel - sol transition point, the decay in $G_e$ and divergence of $\tau_{max,G}$ depends on $\varepsilon$ in a power law fashion expressed as:

$$G_e \sim \varepsilon^z, \quad \text{and} \qquad (15)$$

$$\tau_{max,G} \sim \varepsilon^{-v_G}. \qquad (16)$$

At the critical gel - sol transition state, the system has zero $G_e$ but infinite viscosity. Accordingly, the critical gel neither demonstrates a solid-like behavior, nor exhibits signatures of a liquid state. Beyond the critical transition state, the properties, $\eta_0$ and $\tau_{max,S}$ corresponds to the largest cluster size in the sol state [12]. With increase in temperature beyond the critical temperature, $T_{c,GS}$, as the junction zones dissolve, the largest cluster size decreases. The system shows finite $\eta_0$, characterized by the plateau in the low $\omega$ limit as shown in Figures 2 and 3(a). The Figures 4 (b) and (d) demonstrate that both $\eta_0$ and $\tau_{max,S}$ continue to decrease in a power law fashion with increase in relative distance from critical gel - sol transition state, $\varepsilon$:

$$\eta_0 \sim \varepsilon^{-s}, \quad \text{and} \qquad (17)$$

$$\tau_{max,S} \sim \varepsilon^{-v_S}. \qquad (18)$$

The power law fits the estimated data of $G_e$, $\eta_0$, $\tau_{max,G}$ and $\tau_{max,S}$ represented with Eqs. (15) to (18) are illustrated with solid lines in Figure 4. The scaling exponents can be obtained from these fitting exercises: $z$=2.44, $s$=0.513, $v_G$=2.9, $v_S$=3.05. Consequently, we calculate the critical relaxation exponents in the gel and sol sides using Eqs. (10) and (11), $n_G = z/v_G = 0.828$ and $n_S = 1 - (s/v_S) = 0.835$, respectively. Interestingly, the longest relaxation times show symmetric divergence in the neighborhood of the critical gel state as, $v_S \approx v_G$. This leads to a hyperscaling relationship between the scaling exponents in the sol and gel state, as represented by Eq. (13), $n_H = z/(z+s) = 0.829$. Remarkably, the critical relaxation exponents, $n_G$, $n_S$ and $n_H$ match very well with the critical relaxation exponent, $n_{c,GS} = 0.83$ obtained from the Figure 1 according to Winter- Chambon criteria [36]. Such data, where $n_S \approx n_G$, implies that the critical gel - sol transition state is approached similarly from both the gel and sol state.

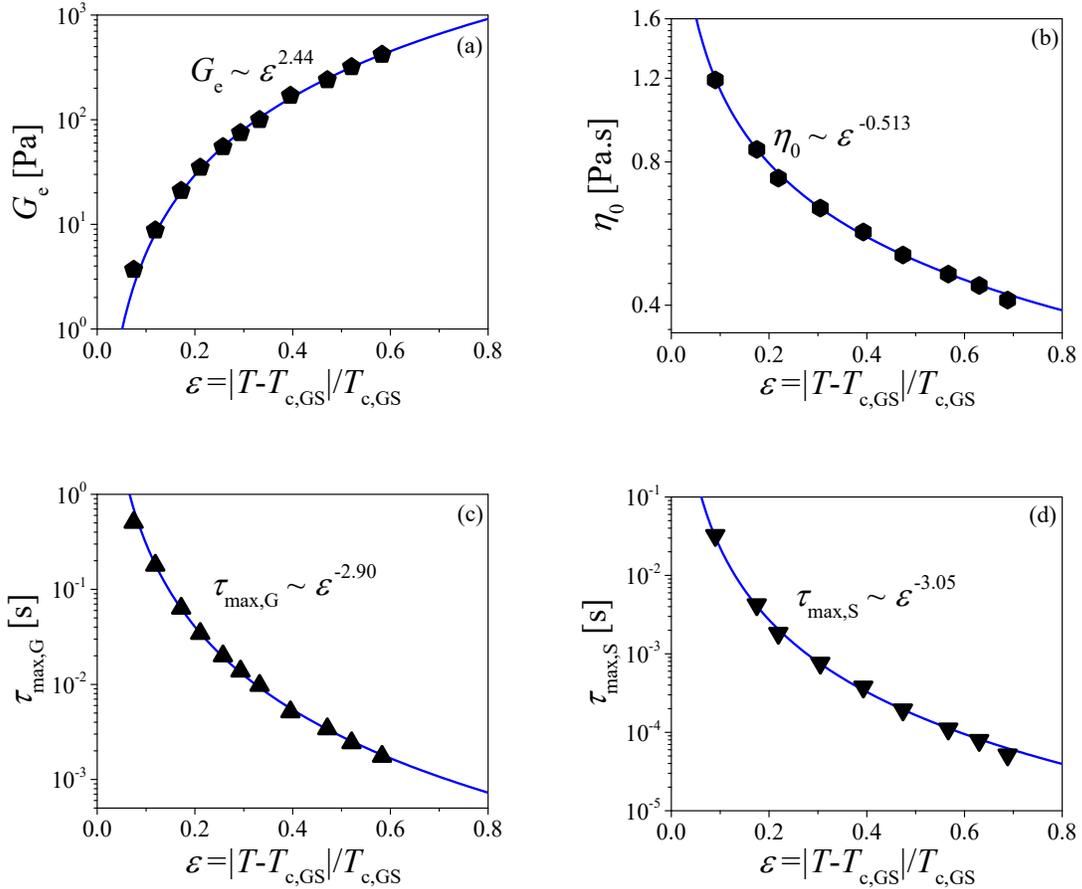

**Figure 4:** Evolution of (a) $G_e$, (b) $\eta_0$, (c) $\tau_{max,G}$ and (d) $\tau_{max,S}$ plotted as a function of relative distance from the critical gel - sol transition state, $\varepsilon = |T - T_{c,GS}|/T_{c,GS}$, where $T_{c,GS}$ is the critical gel - sol transition temperature. The data is shown at a ramp rate of $k = +0.05$ °C/min. The lines show the fitting to the scaling laws expressed through Eqs. (15) to (18) in the gel and sol state.

We investigate the gel to sol transition in the PVOH gel for a range of temperature up ramp rates varying from $k = +0.05$ °C/min to $k = +0.4$ °C/min. It is to be emphasized here that the initial state of the PVOH gel at 0°C is always achieved by an identical procedure of loading the sample at 30°C followed by a temperature ramp down at a fixed ramp rate of $k = -0.05$ °C/min to 0°C. In Figure 5 (a), we plot the critical gel - sol transition temperature, $T_{c,GS}$ (left-hand side ordinate) and the critical gel - sol transition exponent, $n_{c,GS}$ (right-hand side ordinate) as a function of temperature ramp rate, $k$. Interestingly, the gel - sol transition parameter $T_{c,GS}$ and $n_{c,GS}$ is observed to be weakly dependent on the ramp rate $k$. According to Eq. (4), developed by Muthukumar [41], the fractal dimension, $f_{d,GS}$ of the polymeric gel at the critical gel state can be estimated by directly using the relaxation exponent $n_{c,GS}$. The direct relationship between $f_{d,GS}$ and $n_{c,GS}$, leads to $f_{d,GS}$ ranging in the narrow window of 1.48 to 1.53 at all explored temperature ramp rates. This suggests the existence of a similar percolated network structure during heating at all the studied ramp rates. Furthermore, we study the power law dependence associated with the evolution of $G_e$, $\eta_0$, $\tau_{max,G}$ and $\tau_{max,S}$ as a function of the relative distance from the critical gel - sol transition temperature $\varepsilon$ at different temperature ramp rates, $k$. The corresponding scaling exponents associated with $G_e \sim \varepsilon^z$, $\eta_0 \sim \varepsilon^{-s}$ [Eqs. (15) and (17)] are plotted as a function of ramp rate, $k$ in Figure 5 (b). The scaling exponents weakly depend on the ramp rate $k$ with mean values of $z = 2.83 \pm 0.55$, $s = 0.51 \pm 0.01$. On the other hand, the exponent related to the divergence of the longest relaxation times, $v_G$ and $v_S$, [Eqs. (16) and (18)] are plotted in Figure 5(c) as a function of ramp rate, $k$. Within some experimental uncertainty, at all ramp rates, the longest relaxation times show symmetric divergence near the critical gel - sol transition state, $v_G = 3.32 \pm 0.7$, and $v_S = 3.45 \pm 0.4$.

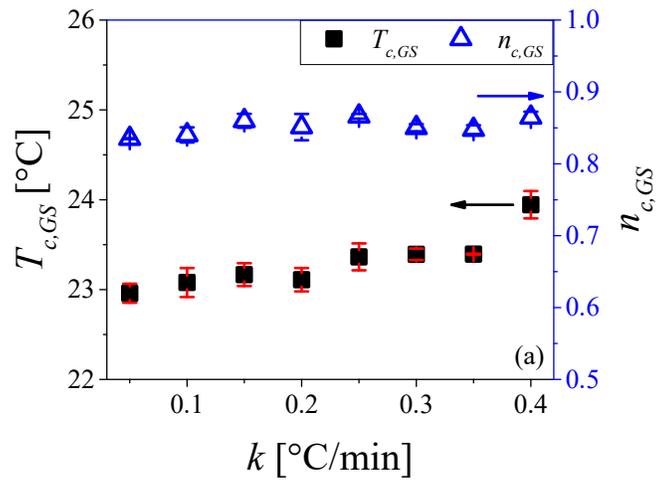

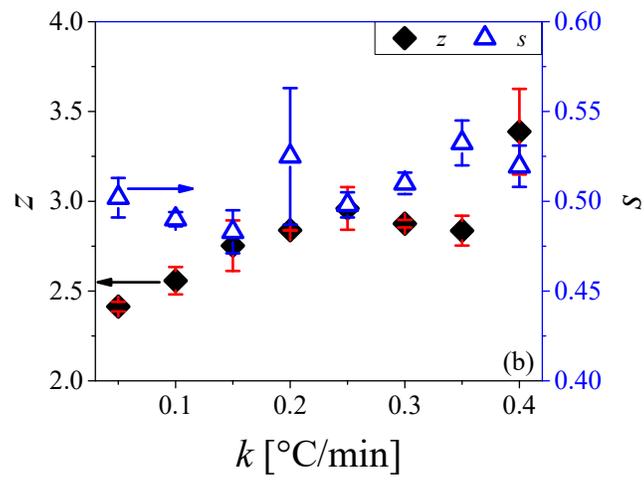

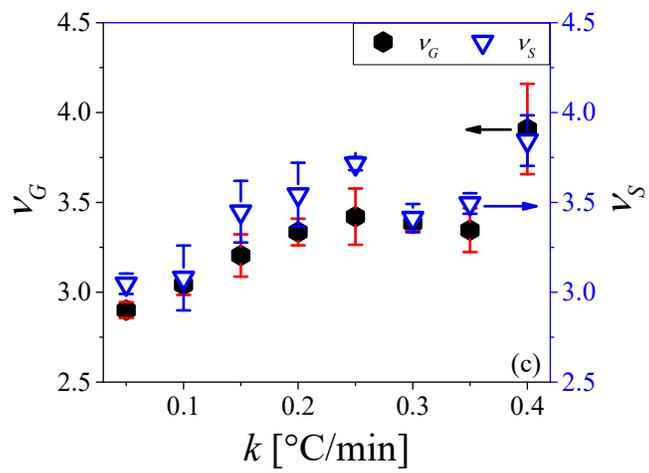

**Figure 5:** (a) The gel to sol transition temperature $T_{c,GS}$ and the critical relaxation exponent $n_{c,GS}$; (b) the scaling exponents $z$ and $s$ associated with $G_e \sim \varepsilon^z$, $\eta_0 \sim \varepsilon^{-s}$; (c) the scaling exponents $v_G$ and $v_S$ associated with $\tau_{max,G} \sim \varepsilon^{-v_G}$, $\tau_{max,S} \sim \varepsilon^{-v_S}$ is plotted as a function of applied temperature ramp rate $k = +0.05\ °C/min$ to $k = +0.4\ °C/min$.

Accordingly, in Figure 6, we replot $|\eta^*|$ normalized with the reference parameter $\eta_{ref}$ as a function of $\omega\tau_{max}$ at different temperatures for all the up ramp rates, $k = +0.05\ °C/min$ to $k = +0.4\ °C/min$, investigated in the present study. As mentioned previously, the normalization parameter for the gel state is $\eta_{ref} = G_e\tau_{max,G}$ and $\tau_{max} = \tau_{max,G}$, while for the sol state is, $\eta_{ref} = \eta_0$ and $\tau_{max} = \tau_{max,S}$. The normalized curves corresponding to all the ramp rates show excellent superpositions independently in the sol and gel domains. The low frequency asymptotes in the gel domain for all ramp rates show an inverse dependence on $\omega$, as represented with the solid line. In the sol domain, the low frequency asymptotes remain independent of $\omega$ and exhibit a constant value of 1, as shown with the dash-dot line. This confirms the observation of the nearly independent values of the power law exponents, $z$, $s$, $v_G$, $v_S$ as a function of up ramp rate $k$ during the gel - sol transition. Furthermore, this also implies that the shape of the relaxation time spectra in the gel and sol state is independent of the heating rate. Consequently, we calculate the critical relaxation exponent associated with the gel side ($n_G$), the sol side ($n_S$) and the hyperscaling relation ($n_H$) adopting the scaling exponents from Figure 5(b) and (c). In Figure 7, we show that the critical relaxation exponents computed from the scaling and hyperscaling relations are in excellent agreement with the one computed using Winter- Chambon criterion [36]: $n_{c,GS} \approx n_S \approx n_G \approx n_H$ for all the ramp rates. Such nearly independent values of the scaling exponents and the critical exponents as a function of ramp rate $k$ further establish a presence of similar kinetics of gelation leading to the percolated network structure. This further suggests that the explored range of heating rates is slow enough to provide an adequate amount of time for the dissociated polymer segments from the network to diffuse and migrate into similar microstructure at the critical state, as evidenced by similar values of $f_d$. However, it might be possible that upon subjecting the system to a very high values of heating rates, the fast change in temperature may allow little rearrangement of the polymer segments and lead to a different microstructure and critical transition temperature. However, such high rates are difficult to explore rheologically, which is constrained by the highest possible frequency range available in a rheometer.

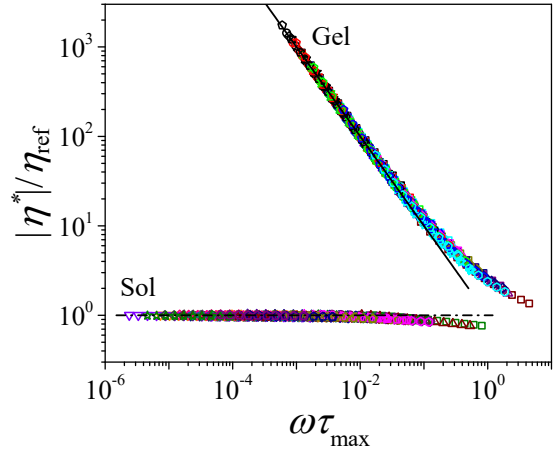

**Figure 6:** Normalized $|\eta^*|$ plotted as a function of normalized $\omega$ at different temperatures for different applied temperature ramp rate $k = +0.05$ °C/min to $k = +0.4$ °C/min during gel - sol transition. The normalization factor for ordinate in the gel state is $\eta_{ref} = G_e \tau_{max,G}$ and in the sol state is $\eta_{ref} = \eta_0$. The normalization factor for the abscissa is the characteristic relaxation time in case of the gel ($\tau_{max,G}$) and the sol ($\tau_{max,S}$) state. The normalized low frequency asymptote in the gel state is shown with a solid line of slope $-1$. The low frequency asymptote in the sol state, shown with a dash-dot line, is the zero-shear viscosity given by: $\eta_0 = \lim_{\omega \to 0} |\eta^*|$.

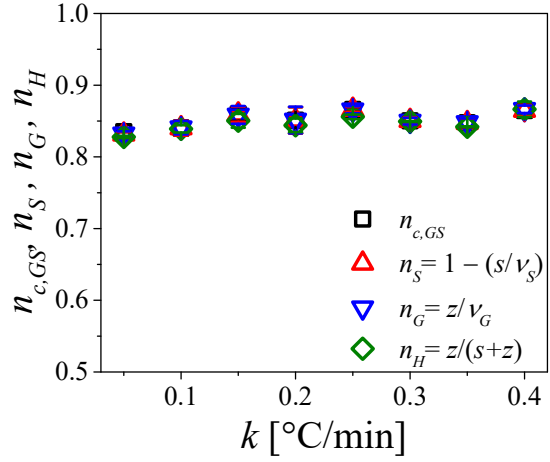

**Figure 7:** The critical relaxation exponent, $n_{c,GS}$ obtained from the isofrequency $\tan\delta$ curves, the critical relaxation exponent obtained from the scaling analysis in the sol state ($n_S$), that of in the gel state ($n_G$) and the critical relaxation exponent obtained from the hyperscaling law ($n_H$) plotted as a function of applied temperature ramp rate $k$.

Scanlan and Winter [13] have experimentally observed that for a system undergoing sol - gel transition through chemical crosslinking, the rate of change of the complex modulus $G^*$ shows a unique power law dependence on $\omega$ with respect to the gelation variable $p$ while approaching the critical sol - gel transition state from the sol side:

$$\left(\frac{\partial \ln G^*}{\partial p}\right)_{p=p_c} \sim \left(\frac{\partial \ln G^*}{\partial t_r}\right)_{t=t_r} \sim \omega^{-\kappa}, \qquad (19)$$

where, $\kappa$ is the dynamic critical exponent and $t_r$ is the reaction time. This relationship suggests that as a sol approaches the critical sol - gel state, the rate of change of the dynamic moduli decreases with an increase in frequency. Consequently, at this state, the other material properties must also be governed by $\kappa$ over all the timescales. Studies have shown that since other scaling relations, Eqs. (15) to (18) hold true for physically crosslinking systems [23, 26, 65, 66], $\Delta p$ in Eq. (19) can be replaced with $\varepsilon$, such that in the close to vicinity to critical sol - gel transition state we have,

$$\left(\frac{\partial \ln G^*}{\partial p}\right)_{p=p_c} \sim \left(\frac{\partial \ln G^*}{\partial X}\right)_{X=X_c} \sim \omega^{-\kappa}, \tag{20}$$

where, $X = t$ for spontaneously evolving systems and $X = T$ for temperature dependent gel forming systems [23, 26, 39]. Furthermore, Eq. (20) can be represented in terms of $G'$ and $G''$ as:

$$\left(\frac{\partial \ln G'}{\partial X}\right)_{X=X_c} = C\left(\frac{\partial \ln G''}{\partial X}\right)_{X=X_c} \sim \omega^{-\kappa}, \tag{21}$$

where, $C$ is the constant of proportionality. In the literature, the values of $\kappa$ and $C$ are obtained both theoretically [20, 77] and experimentally [13, 23, 26, 39]. Scanlan and Winter [13] have reported $\kappa = 0.21 \pm 0.02$ and $C \approx 2$ for over 30 experiments on sol - gel transition of PDMS through chemical crosslinking with different stoichiometry, chain lengths and concentration. Joshi and coworkers [23, 39], for over 50 experiments carried out on a spontaneously evolving hectorite nanoclay with various concentrations, computed $\kappa \approx 0.2$ and $C \approx 2$. For an aqueous PVOH solution undergoing sol - gel transition with decrease in temperature, $\kappa \approx 0.22$ and $C \approx 2.3$ have been reported [23]. Such reports, where $\kappa \approx 0.2$ and $C \approx 2$, emphasize the underlying similarity of the process of sol - gel transition irrespective of the nature of crosslinking and mechanism of gelation.

In the present case of gel - sol transition, we represent Eq. (21) in terms of temperature, $T$ while the derivative is computed at $X_c = T_{c,GS}$. The derivative is computed in two ways: one being calculated from the gel side using the forward difference method and the second computed from the sol side using the backward difference method. Therefore, the parameters $\kappa_E$ and $C_E$ (suffix -$E$ represents experimental measurement), are obtained from the interrelations of the experimental power law dependence of the rate of change of $G^*$, $G'$ and $G''$ on $\omega$ at the critical gel state from the gel side (represented with suffix $G$) and the sol side (represented with suffix $S$). The parameter $\kappa_{E,G}$ and $\kappa_{E,S}$ reported in this study is the average of the power law exponent observed for the rate of change of $G'$ and $G''$ on $\omega$. The corresponding power law exponents $\kappa_{E,G}$ and $\kappa_{E,S}$ and the constants of proportionality $C_{E,G}$ and $C_{E,S}$ are plotted as a function of the ramp rate of increasing temperature, $k$ in Figure 8. We observe that for all values $k$, $\kappa_{E,G} = \kappa_{E,S} = 0.26 \pm 0.03$ while $C_{E,G} = C_{E,S} = 2.9 \pm 0.1$. Such identical values of $\kappa_E$ and $C_E$, when computed from the gel and the sol sides, suggests that these values remain consistent in the neighborhood of the critical gel state. The value $C_E \approx 2.9$ suggests that the rate of change of $G'$ is 2.9 times faster than the rate of change of $G''$ during the gel to sol transition. The computed values of $\kappa$ and $C$ suggest a similarity in the process of the gel - sol transition with sol - gel transition for a thermoresponsive system. To the best of our knowledge, this work is the first report to quantify the value of $\kappa$ and $C$ during gel - sol transition.

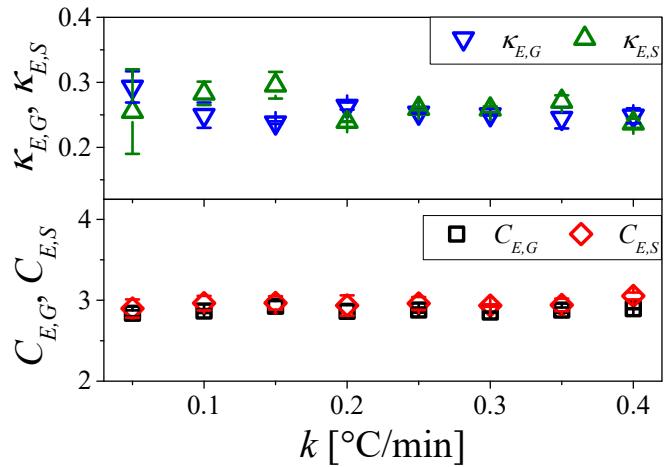

**Figure 8:** The power law exponent ($\kappa_E$) and the constant of proportionality ($C_E$) associated with dependence of the rate of decrease in $G^*$, $G'$ and $G''$ on $\omega$ according to Eqs. (20) and (21) are plotted as a function of the applied temperature ramp rate $k$. The reported values of the parameters, $\kappa_{E,G}$ and $\kappa_{E,S}$, are the average of the power law exponent observed for the rate of change of $G'$ and $G''$ on $\omega$. The values measured during the evolution from the gel to critical gel - sol transition state are represented with a suffix $G$, and the evolution from the critical gel - sol transition state to the sol is represented with the suffix $S$.

For several physically crosslinked systems such as aqueous dispersions of nanoclay, chitosan hydrogels, sol - gel transition of PVOH etc., the equivalence of $\Delta p \sim \varepsilon$ has been confirmed in the literature [23, 26, 65, 66]. Negi and coworkers [26] studied a spontaneously evolving system of colloidal silica particles with octadecyl chains dispersed in decalin. They obtained the scaling exponents $s$ and $z$ from the direct experimental measurement of $\kappa$ and $n_c$:

$$\kappa = (1 - n_c)/s, \quad \text{and} \tag{22}$$

$$\kappa = n_c/z. \tag{23}$$

Strictly speaking, the relations given by Eqs. (22) and (23) between $\kappa$, $n_c$, and $z$, are simultaneously valid only for the special case of symmetric divergence of the longest relaxation time. Interestingly, this formulation also leads to a relation between $v_S$, $v_G$ and $\kappa$ as: $v_S^{-1} = v_G^{-1} = \kappa$, suggesting that the relaxation time diverges symmetrically in both sides of the critical state and is directly dependent on the growth rate of the moduli through the exponent $\kappa$ [23]. Based on the theoretical range of $0 < n_c < 1$ the above equations relating $\kappa$, $n_c$, $s$ and $z$ also define the range of $s$ and $z$ as $0 < s, z < 1/\kappa$. Furthermore, Eq. (22) and (23) can mathematically lead to a single hyperscaling relation:

$$\kappa = 1/(s + z). \tag{24}$$

This suggests that the exponents $s$ and $z$ cannot take independent values between 0 to $1/\kappa$, and the sum must be equal to $1/\kappa$. Accordingly, the experimental reports on $\kappa \approx 0.2$ imply that $s + z \approx 5$. Suman and Joshi [23] showed excellent agreement between the values of $\kappa$ obtained from experimental measurement with the predictions from the Eq. (22) for over 20 experiments associated with sol - gel transition carried out on colloidal and polymeric gels. Figure 9 compares the $\kappa_E$ values measured from the gel - sol transition experiments through Eq. (21) with the $\kappa_A$ values predicted from the scaling analysis (represented with suffix A) following Eq. (22) for the gel - sol transition. The power law exponent $\kappa_{E,G}$ and $\kappa_{E,S}$ determined from the oscillatory shear measurements matches quite well with $\kappa_A$ computed from the scaling analysis [Eq. (22)] within the experimental uncertainty. According to the percolation theory analysis of de Gennes [20], the value of $z$ can be further related to the power law exponent corresponding to the divergence of cluster size, $\alpha$ as: $z = 1 + (d - 2)\alpha$, where $d$ is the dimension of space. With a mean value of $z = 2.8$ for PVOH gel, we get $\alpha = 1.8$. This suggests that the largest cluster size ($\xi$) decays as $\xi \sim \varepsilon^{-1.8}$ near the critical transition state.

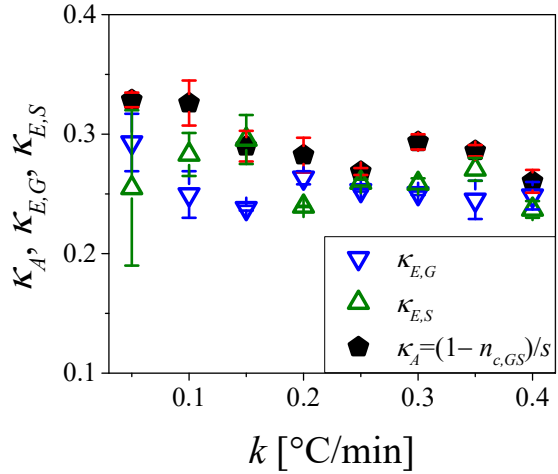

**Figure 9:** The power law exponent, $\kappa_{E,G}$ & $\kappa_{E,S}$ obtained from analysis and $\kappa_A = (1 - n_{c,GS})/s$ is plotted as a function of the applied temperature ramp rate $k$. The values measured during the evolution from the gel to critical gel - sol transition state are represented with the suffix $G$, and the evolution from the critical gel - sol transition state to the sol is represented with the suffix $S$.

In this context, it is important to elaborate on the distinct features of the gel - sol transition. Firstly, the critical sol - gel transition temperature ($T_{c,SG} = 8.6°C$, as shown in Figure S1 of the supplementary information) of aqueous PVOH solution is much lower compared to the critical gel - sol transition temperature ($T_{c,GS} = 23.1°C$). This suggests that the three-dimensional percolated network structure of the gel is conserved up to a much higher values of temperature before it undergoes a gel – sol transition. Additionally, the nature of the network structure is also different at the critical gel - sol transition state as compared to sol - gel transition state (refer to Figure S1 of the supplementary information), as $n_{c,GS} < n_{c,SG}$. This difference in the critical relaxation exponents, $n_{c,SG} = 0.93$ ($f_{d,SG} = 1.38$) and $n_{c,GS} = 0.83$ ($f_{d,GS} = 1.54$), suggests that percolated network structure is denser during the gel - sol transition. Consequently, although the hyperscaling relations are valid for both sol - gel and gel - sol transition, the scaling exponents $z$, $s$, $\upsilon_G$, and $\upsilon_S$ may not be directly related to each other for the two transition processes. Secondly, for a variety of sol - gel transition processes in gelatin methacryloyl [29], Carrageenan [78], Pluronic F108 [79], the network structure formation has been reported to demonstrate a strong dependence on the applied temperature ramp rate. On the contrary, for the gel - sol transition, we observe the critical gel - sol transition temperature, $T_{c,GS}$, and the critical relaxation exponent, $n_{c,GS}$ to be nearly independent of temperature ramp rate over the investigated range of $k = +0.05$ °C/min to $k = +0.4$ °C/min. For a gelatin methacryloyl system, Park et al. [29] have also reported gel - sol transition temperature to remain independent of heating ramp rate varying from 0.3 °C/min to 10 °C/min. Interestingly, our results match very well with this trend.

The validation of the hyperscaling relations Eq. (6) to (13) for gel - sol transition in Figure 4 confirms the consistency of the relation $\varepsilon = |T - T_{c,GS}|/T_{c,GS}$ as the measure of the extent of reaction, $\Delta \tilde{p}$ for the gel - sol transition in a physical gel. Such equivalence of $\Delta \tilde{p} \sim \varepsilon$ has been reported to hold true only when the rate of formation of crosslinking is a first order with respect to reacting units [14]. This suggests that the gel - sol transition is a first order reaction with the variable $T$. The equivalence of $\Delta \tilde{p} \sim \varepsilon$ also validates the computation of $\kappa_{E,G}$, $\kappa_{E,S}$ and $C_{E,G}$, $C_{E,S}$ using the Eq. (21). This provides two fundamental insights into the gel - sol transition. Firstly, we obtain the power law dependence of the rate of decay of dynamic moduli on $\omega$ during the gel - sol transition, $\kappa_{E,G} \approx \kappa_{E,S} \approx 0.26$. This closely agrees to the value of the dynamic critical exponent, $\kappa \approx 0.2$, experimentally measured for several physical and chemical gels. Secondly, for the gel - sol transition, the experimentally measured value of $C_{E,G} \approx C_{E,S} \approx 2.9$ closely resembles to $C \approx 2$ reported for sol - gel transition. This suggests that while the growth in $G'$ is twice faster than $G''$ during sol - gel transition, the decay in $G'$ is 2.9 times faster than $G''$ during gel - sol transition. Furthermore, we compute the values of $\kappa_A$, using the scaling exponents $n_{c,GS}$, $z$ and $s$ according to Eq. (22) and (23). Figure 9 shows that the analytically computed values of $\kappa_A$ show good agreement with the experimental measurements. This corroborates the universality of the hyperscaling relations, relating $n_{c,GS}$, $z$, $s$, $v_G$, $v_S$ and $\kappa_E$, proposed for a sol - gel transition to a gel - sol transition and affirms the usage of $\Delta \tilde{p} \sim \varepsilon = |T - T_{c,GS}|/T_{c,GS}$. The analogous nature of the evolution of viscoelastic properties highlights the similarities of the two opposite processes, sol - gel and gel - sol transition. This work highlights the applicability of the scaling relations to any crosslinking system undergoing gel - sol transition and ascertains the universality of the scaling relations. Furthermore, the validation of the scaling relations in a system undergoing gel - sol transition highlights the similarity in the qualitative evolution of the viscoelastic properties as the critical gel state is achieved from the sol or gel side. With the validation of scaling relations, it becomes very convenient to have a priori knowledge of the viscoelastic behavior of the material at any time during the gel - sol transition.

**Conclusions**

The present work investigates the applicability of the scaling theories, originally proposed for sol – gel transition, on the gel - sol transition of a thermoresponsive PVOH gel. On decreasing the temperature of an aqueous solution of PVOH, the system undergoes sol - gel transition owing to network structure formation through hydrogen bonding. The hydrogen bonds break on reversal of the temperature field, and consequently, the system undergoes a gel - sol transition. On performing oscillatory shear measurements, gel - sol transition demonstrates a critical transition state where the material properties are scale invariant, suggesting the existence of the weakest self-similar space spanning percolated network structure. We study the evolution of the linear viscoelastic properties during the gel - sol transition. We utilize the complex viscosity data to evaluate the equilibrium modulus, $G_e$, zero shear viscosity, $\eta_0$ and the longest relaxation times in the gel ($\tau_{max,G}$) and the sol ($\tau_{max,S}$) state. The viscoelastic properties in the gel state, $G_e$ and $\tau_{max,G}$ follow a power law dependence according to Eqs. (15) and (16), where, $\varepsilon = |T - T_{c,GS}|/T_{c,GS}$ is expressed according to previous reports [23] on temperature dependent sol - gel transition of polymeric gels. Similarly, in the sol state, $\eta_0$ and $\tau_{max,S}$ decays with increase in temperature according to Eqs. (17) and (19). Interestingly, the critical relaxation exponent computed from the scaling exponents in the gel state ($n_G$), sol state ($n_S$) and from the hyperscaling relationship ($n_H$) precisely agrees with the critical relaxation exponent ($n_{c,GS}$) obtained experimentally using the Winter criterion from the $\tan\delta$ curve. Furthermore, near the critical gel - sol transition state, as the dynamic moduli decreases, we experimentally measure the power law dependence associated with the rate of decrease in dynamic moduli with angular frequency, $\omega$ as, $\kappa_{E,G} \approx \kappa_{E,S} \approx 0.26$. Interestingly, this value of $\kappa_{E,G}$ and $\kappa_{E,S}$ closely relates to the dynamic critical exponent, $\kappa \sim 0.2$, reported for the sol - gel transition of various chemical and physical gels [13, 23, 39]. The experimentally observed value of $\kappa_{E,G}$ and $\kappa_{E,S}$ also agrees with the predictions from the scaling exponents. This work, therefore, comprehensively confirms the applicability of the hyperscaling relations for the gel - sol transition. While the sol - gel and gel - sol transition processes are opposite to each other, the applicability of the hyperscaling relations highlights the similarity in the evolution of the material properties in the process of gel - sol transition with the sol - gel transition.

**List of symbols**

| | |
|---|---|
| $C$ | Constant of proportionality relating rate of change of dynamic moduli [-] |
| $C_{E,G}$ | Constant of proportionality calculated from gel side [-] |
| $C_{E,S}$ | Constant of proportionality calculated to sol side [-] |
| $d$ | Dimension of space |
| $f_d$ | Fractal dimension |
| $f_{d,GS}$ | Fractal dimension associated to gel - sol transition |
| $G(t)$ | Relaxation modulus [Pa] |
| $G_0$ | Modulus associated with fully developed network structure [Pa] |
| $G^*$ | Complex modulus [Pa] |
| $G'$ | Elastic modulus [Pa] |
| $G''$ | Viscous modulus [Pa] |
| $G_e$ | Equilibrium modulus [Pa] |
| $H(\tau)$ | Continuous relaxation time spectra [Pa] |
| $k$ | Cooling/ Heating ramp rate [°C/$min$] |
| $n_c$ | Critical relaxation exponent [-] |
| $n_{c,SG}$ | Critical relaxation exponent associated to sol - gel transition [-] |
| $n_{c,GS}$ | Critical relaxation exponent associated to gel - sol transition [-] |
| $n_S$ | Critical relaxation exponent calculated from sol side [-] |
| $n_G$ | Critical relaxation exponent calculated from gel side [-] |
| $n_H$ | Critical relaxation exponent associated to hyperscaling relationship [-] |
| $p$ | Extent of network connectivity [-] |

| | | |
|---|---|---|
| $\Delta \tilde{p}$ | Normalized distance from critical gel state in terms of connectivity [-] |
| $s$ | Power law exponent associated to zero shear viscosity [-] |
| $S$ | Gel strength [Pa.$s^{n_c}$] |
| $t$ | Time [s] |
| $t_r$ | Reaction time [s] |
| $T$ | Temperature [°C] |
| $T_{c,SG}$ | Critical sol - gel transition temperature [°C] |
| $T_{c,GS}$ | Critical gel - sol transition temperature [°C] |
| $z$ | Power law exponent associated to equilibrium modulus [-] |

**Greek Symbols:**

| | |
|---|---|
| $\alpha$ | Power law exponent associated with growth/ decay of largest cluster [-] |
| $\varepsilon$ | Normalized distance from critical gel state to sol or gel state [-] |
| $\eta^*$ | Complex viscosity [Pa.s] |
| $\eta_0$ | Zero shear viscosity [Pa.s] |
| $\kappa$ | Dynamic critical exponent [-] |
| $\kappa_{E,G}$ | Dynamic critical exponent calculated from gel side [-] |
| $\kappa_{E,S}$ | Dynamic critical exponent calculated to sol side [-] |
| $\nu_S$ | Power law exponent associated to longest relaxation time in sol state [-] |
| $\nu_G$ | Power law exponent associated to longest relaxation time in gel state [-] |
| $\xi$ | Largest cluster size [lengthscale] |
| $\tau$ | Relaxation time [s] |
| $\tau_0$ | Relaxation time associated with the primitive link [s] |
| $\tau_{max}$ | Longest relaxation time in sol/ pre-gel or gel/ post-gel state [s] |
| $\tau_{max,S}$ | Longest relaxation time in sol/ pre-gel state [s] |
| $\tau_{max,G}$ | Longest relaxation time in gel/ post-gel state [s] |
| $\omega$ | Angular frequency [rad/s] |

**Appendix**

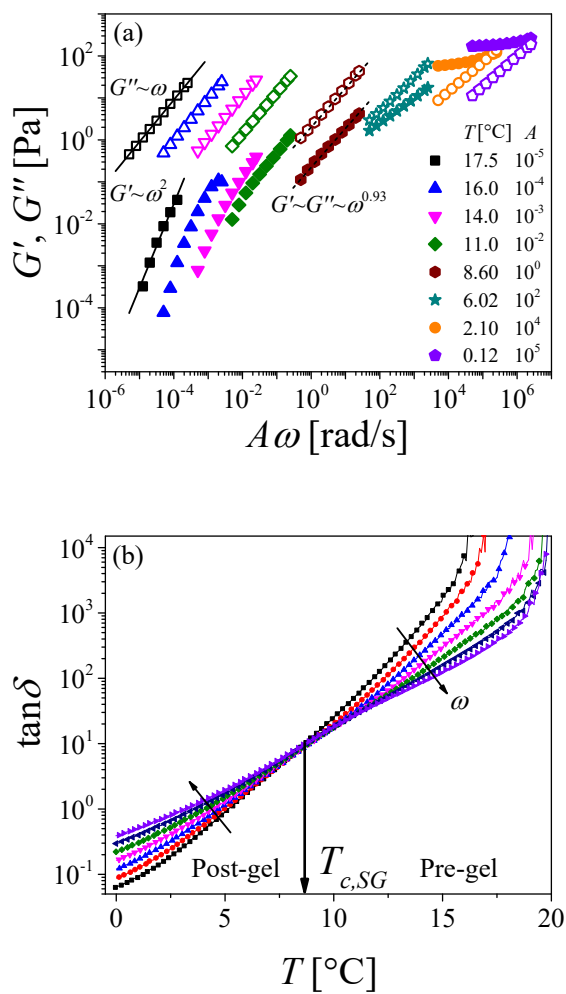

**Figure A1:** (a) Evolution of $G'$ (closed symbol) and $G''$ (open symbol) plotted as a function of $\omega$ at different temperatures on the application of a temperature ramp at a rate of $k = -0.05$ °C/min along with cyclic frequency sweep. The data has been shifted horizontally with a shift factor $A$ for better clarity, and the magnitude of $A$ at any temperature is mentioned in the legends. The dependence of $G'$ and $G''$ on $\omega$ in the pre-gel state ($G' \sim \omega^2$ and $G'' \sim \omega$) and the critical sol - gel transition state ($G' \sim G'' \sim \omega^{0.93}$) is shown with a solid line and a dashed line respectively. (b) The corresponding evolution of $\tan\delta$ as a function of temperature is shown for the sol-gel transition. At the critical sol - gel transition state, where $\tan\delta$ is independent of $\omega$, $T_{c,SG}$ is marked with a solid arrow. The slanted arrows show the direction of increasing $\omega$.

The freshly prepared 12 wt.% aqueous solution of Poly(vinyl alcohol) (PVOH) is cooled from 30 °C to 0 °C at a ramp rate of $k = -0.05$ °C/min. During cooling, it is subjected to cyclic frequency sweep measurements in the range of angular frequency, $\omega = 0.5 - 25$ rad/s. In Figure S1, we illustrate the corresponding evolution of elastic ($G'$) and viscous ($G''$) moduli and $\tan \delta$ as a function of angular frequency, $\omega$ as well as temperature. The Figure S1 (a) presents the dependence of $G'$ and $G''$ on $\omega$ at different temperatures. The horizontal axis of $\omega$ has been shifted with a shift factor $A$ for clarity. In the pre-gel state, the system shows signatures of liquid like behavior at higher temperatures, where $G' \sim \omega^2$ and $G'' \sim \omega$, as shown in the figure with a solid line. The corresponding change in $\tan \delta$ is shown in figure S1 (b), where $\tan \delta \gg 1$ (as $G'' \gg G'$) and decreases with an increase in $\omega$ in the liquid state. With further decrease in temperature, the dependence of $G'$ and $G''$ on $\omega$ gradually weakens, more specifically for $G'$ at a greater intensity. At the critical sol - gel transition temperature, $T_{c,SG} = 8.6$ °C, both $G'$ and $G''$ show identical power law dependence on $\omega$ with critical relaxation exponent, $n_{c,SG} = 0.93$, as shown in the figure with a dashed line. Accordingly, the isofrequency $\tan \delta$ curves merge at a single point, as shown by the solid arrow. As the gel consolidates with further decrease in temperature, $G'$ more strongly increases compared to $G''$ and $\tan \delta$ decreases. In the post-gel state, as $G'$ shows a plateau and $G''$ exhibits a weak dependence on $\omega$, $\tan \delta$ increases with an increase in $\omega$.

In previous studies from our group [Joshi et al., Macromolecules 53 (2020) 3452, Suman and Joshi, Journal of Rheology 64 (2020) 863], the scaling laws associated with the sol-gel transition of aqueous PVOH solution on cooling [Eq. (6) to (13)] have been verified extensively for different molecular weights of PVOH as well as a wide range of cooling rates. The present study focuses on the applicability of the scaling relations for the gel - sol transition of an aqueous PVOH system on heating.

**Acknowledgements:** YMJ acknowledges financial support from the Science and Engineering Research Board, Government of India (Grant number: CRG/2018/003861).

**Data Availability Statement:** The experimental data that supports the findings in the present is available from the corresponding authors on reasonable request.